\documentclass[a4paper, conference]{IEEEtran}
\IEEEoverridecommandlockouts
\usepackage{amsmath,amssymb,amsfonts}
\usepackage{algorithmic}
\usepackage{graphicx}
\usepackage{textcomp}
\usepackage{xcolor}
\usepackage[square,numbers,sort]{natbib}
\usepackage[obeyspaces]{xurl}
\usepackage[T1]{fontenc}
\usepackage{tikz}
\usetikzlibrary{external}
\usepackage{multirow}
\usepackage{enumitem}

\usepackage[compact]{titlesec}
\titlespacing{\section}{0pt}{5pt}{2pt}
\titlespacing{\subsection}{0pt}{2pt}{2pt}
\titlespacing{\subsubsection}{0pt}{2pt}{2pt}
\IEEEoverridecommandlockouts
\IEEEpubid{\makebox[\columnwidth]{ 978-1-6654-4399-9/21/\$31.00
\copyright~2021~IEEE }
\hspace{\columnsep}\makebox[\columnwidth]{ }}

\begin{document}

\title{TypoSwype: An Imaging Approach to Detect Typo-Squatting\\
{\footnotesize \textsuperscript{}}
\thanks{}
\vspace{-30pt}

}

\author{\IEEEauthorblockN{Lee Joon Sern}
\IEEEauthorblockA{\textit{Ensign Labs} \\
\textit{Ensign InfoSecurity}\\
Singapore \\
lee\_joonsern@ensigninfosecurity.com}
\and
\IEEEauthorblockN{Yam Gui Peng David}
\IEEEauthorblockA{\textit{Ensign Labs} \\
\textit{Ensign InfoSecurity}\\
Singapore \\
david\_yam@ensigninfosecurity.com}
}

\IEEEaftertitletext{\vspace{-30pt}}
\maketitle

\begin{abstract}
    Typo-squatting domains are a common cyber-attack technique. It involves utilising domain names, that exploit possible typographical errors of commonly visited domains, to carry out malicious activities such as phishing, malware installation, etc. Current approaches typically revolve around string comparison algorithms like the Demaru-Levenschtein Distance (DLD) algorithm. Such techniques do not take into account keyboard distance, which researchers find to have a strong correlation with typical typographical errors and are trying to take account of. In this paper, we present the TypoSwype framework which converts strings to images that take into account keyboard location innately. We also show how modern state of the art image recognition techniques involving Convolutional Neural Networks, trained via either Triplet Loss or NT-Xent Loss, can be applied to learn a mapping to a lower dimensional space where distances correspond to image, and equivalently, textual similarity. Finally, we also demonstrate our method's ability to improve typo-squatting detection over the widely used DLD algorithm, while maintaining the classification accuracy as to which domain the input domain was attempting to typo-squat.
\end{abstract}

\begin{IEEEkeywords}
    Computer Network security, Neural Networks
\end{IEEEkeywords}

\section{Introduction}
    A common type of cyber attack involves registering domains which utilize typographical errors that unsuspecting users may commit when keying in desired domains. These are commonly known as typo-squatting domains. In 2006, a typo-squatted variant of "google[.]com", "goggle[.]com" was abused by adware. Recently, a typo-squatted variant of "youtube[.]com", "yuube[.]com" was used to host malware \cite{wikiTypo}.
    
    For typo-squatting, the current industry accepted approach is to calculate the edit distance between strings. Equation \eqref{eq1} shows two possible domains which are 1 Levenshtein edit Distance (LD) away from "facebook.com". In this method, a lower LD value indicates more similar domains, increasing the confidence of a phishing attempt.
    \begin{equation}
    \begin{split}
        ld("facebook.com", "fapebook.com") = 1\\
        ld("facebook.com", "faceb0ok.com") = 1 \label{eq1}
    \end{split}
    \end{equation}
    
    However, this method fails to account for keyboard distance. For example, in Equation \eqref{eq1}, "faceb0ok.com" is more likely to be a typo-squatted variant of "facebook.com" because "o" and "0" are very close together while "p" and "c" are extremely far apart. Despite this stark difference, the LD algorithm classifies the 2 to be of equal edit distance from "facebook.com".
    
    Notwithstanding this, the LD still forms the basis of many modern day spell checking systems, making it suitable for detecting typo-squatting domains, which exploit typographical/ spelling errors. LD consists of 3 atomic operations; insertion, deletion and substitution. The LD between strings is defined as the smallest number of such operations required to transform one string to another \cite{Leven}. Building upon this work, the Dameru-Levenschtein Distance (DLD) algorithm was created to account for an additional operation; transposition of adjacent characters \cite{Dameru}. The DLD algorithm is still being widely used today and is the core algorithm used in both Google and Spotify to correct user's spelling error \cite{Samuelsson}.
    
    In recent times, researchers have noted the importance of taking into account keyboard distance and are increasingly finding new ways to account for it and come up with better spell check algorithms. \citeauthor{stringkeyboard} proposed a simplistic way of comparing 2 strings by returning the sum of distances between corresponding characters within. If one string is longer than the other, the remaining characters are counted as having the same value as the maximum distance. Although it takes into account keyboard distance, it fails to take into the various actions that DLD-based algorithms account for. 
    
    \citeauthor{Samuelsson} \cite{Samuelsson} proposed infusing some notion of keyboard distance into the DLD algorithm. He proposed weighting deletion and substitution operations based on keyboard layouts, while leaving insertion and transposition operations untouched. He found that given his large Spotify dataset consisting of more than a million string pairs of spelling errors and their targets, there was a large correlation between the physical key locations on the keyboard and spelling mistakes , affirming the importance of keyboard distance in spell checks. However, \citeauthor{Samuelsson} was unable to demonstrate any significant advantage over the baseline DLD algorithm. Another drawback noted by \citeauthor{Samuelsson} is the need to define weights for deletion and substitution algorithms. Unlike the DLD algorithm where this is a widely accepted set of weights to use for each of the 4 operations, \citeauthor{Samuelsson} noted it was challenging to come up with standard weights that can generalise across datasets.
    
    \citeauthor{moubayed2020ensemble} \cite{moubayed2020ensemble} showed that Machine Learning methods could reduce the computational complexity associated with dynamic programming approaches like the DLD algorithm, while maintaining accuracy and precision, supporting the use of learning-based models for typo-squatting.
    
    Recently, \citeauthor{Movin} showed that it was possible to make use of RNNs to correct spelling errors \cite{Movin}. However, \citeauthor{Movin} noted that there was a problem deploying such a system as the RNN model was configured to output a series of characters. There is no guarantee that the output series of characters is part of a desired vocabulary of words. Furthermore, the series of characters output by the RNN model will also be biased to the vocabulary of words that was used to train it.
    
    We now highlight our main contributions. Our major contribution is TypoSwype, a novel approach to detect typo-squatting domains through the use of the Swype framework developed by Swype Inc. Swype is a virtual keyboard for touchscreen smartphones and tablets where users enter words by sliding a finger or stylus from the first to last letter of a word  \cite{wikiSwype}. Our proposed TypoSwype framework first renders an image, $Q$, by tracing lines that would be formed when sliding a particular queried domain name, over a keyboard, from the first to the last character just like how Swype works. 
    
    Our next contribution is the demonstration of how state of the art (SOTA) image recognition algorithms, based on Convolutional Neural Networks (CNN), can be used to determine if the generated image is similar to a database of images, $R = \{r_1,r_2,r_3,...,r_N\}$, generated from the top $N$ popular domain names on the Internet. If $Q$ is deemed to be similar to any one of the images in the set of $R$, the queried domain name corresponding to $Q$ is deemed to be a typo-squatted domain of the domain name whose corresponding image, in the set of $R$, $Q$ is deemed to be most similar to. It should be noted that Swype itself has an auto-complete algorithm which determines words based on how the user slides over the virtual keyboard, however, its primary purpose is for auto-completion and not spell checking \cite{swype_work}. In particular, it assumes that the input path is correct for it to guess a word. To the best of our knowledge, this is the first of its kind application of CNNs. An advantage of this approach over \citeauthor{Movin}'s work is that the output will be part of a known, accepted vocabulary. There is no risk that the model generates a word that isn't part of the pre-defined vocabulary.
    
\section{Preliminaries and Problem Setup}
    In this section, we review how malware binaries can be converted to images.

    In this section, we briefly review the current state of the art image recognition algorithms that are able to learn similarity functions that measure the similarity between a pair of objects.
    
    As mentioned earlier, the crux of our TypoSwype framework is the conversion of the spell-checking problem to the image domain via swype-like images, which takes into account keyboard layouts innately. Spell-checking can then be done by determining similarity between the input image and a database of images, corresponding to a given checking list of domains (or equivalently a given vocabulary of common words). Thus, we require an image recognition algorithm that can learn an encoding function, $E(i) = e$, where $E$ is the encoding function, $i$ is the input image and $e$ is a summarized embedding of the input image, $i$. The encoding function should output similar embeddings for similar swype-like images, indicating the possibility of a typo-squatting domain, and dissimilar embeddings for dissimilar swype-like images, indicating a benign domain. To solve this problem, we turn to SOTA algorithms, the Triplet loss (TL) and NT-Xent loss (NL). 
    
\begin{figure}[!htp]
        \centering
        \includegraphics[width=5cm]{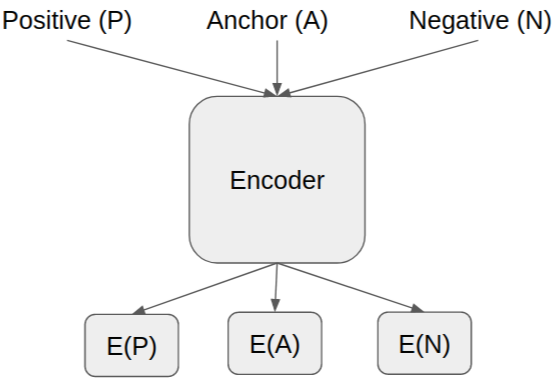}
        \caption{Illustration of Triplet Loss}
        \label{fig:triplet_loss_eg}
    \end{figure}

    Fig \ref{fig:triplet_loss_eg} illustrates how TL works. It comprises an Encoder configured to take in 3 input images; positive (P), anchor (A) and negative (N) images, during the training process. The positive image is the image that is similar to the anchor image while the negative image is an image that is dissimilar to the anchor image. The Encoder which typically comprises a deep CNN is configured to output embeddings (i.e. a vector representation) for each input image. Finally the Encoder can be trained via the loss function \eqref{triplet_loss}. Minimising $L_{triplet}$ is equivalent to minimising the distance between $E(A)$ and $E(P)$ such that it is at least smaller than the distance between $E(A)$ and $E(N)$ with a margin of $M$. The TL is an improvement over the pair loss used in Siamese Networks and has been demonstrated by \citeauthor{facenet} to be able to train a CNN to achieve an accuracy of 99.63\% on the Learning Faces in the Wild dataset consisting of more than 50k faces.
    
    \small
    \begin{equation}
        L_{triplet} = max(||E(A)-E(P)||^2  - ||E(A)-E(N)||^2 + M, 0) \label{triplet_loss}
    \end{equation}
    \normalsize
    
    One drawback of the TL is the difficulty to select the negative samples. \citeauthor{facenet} noted that training on the hardest pairs (i.e. the pairs where the model makes the largest errors) typically resulted in a bad local minima early on in the training \cite{facenet}. Such phenomenon was also observed by \citeauthor{xuan2020hard} where they proposed modifying the loss function in order to make use of the hardest pairs for stable training \cite{xuan2020hard}.
    
    Recently \citeauthor{simclr} introduced the NL loss function \eqref{NL_loss}, which can be used to train the Encoder \cite{simclr}. It is similar to the softmax function but adjusted with a temperature term, $\tau$. For each anchor positive pair, it takes in a batch of negative samples, $b_n$, and using the similarity function defined in \eqref{NL_sim_func}, which is essentially cosine similarity, the NL loss, $L_{NT-Xent}$, can be computed. Gradients can subsequently be back-propagated to update the Encoder. Note that this formulation precludes the need to choose triplets. Instead it selects a batch of negative samples for each anchor-positive pair, thus circumventing the problem of unstable training when selecting the hardest pairs. \citeauthor{simclr} showed that they were able to achieve state of the art performance on the ImageNet dataset when comparing to prior semi-supervised and self-supervised work.
    
    \small
    \begin{equation}
        s_{i,j} = \frac{s_i^Ts_j}{||s_i||||s_j||}
        \label{NL_sim_func}
    \end{equation}
    \normalsize
    
    \small
    \begin{equation}
        L_{NT-Xent} = -log \left[\frac{e^{s_{a,p}/\tau}} {\sum_{n=1}^{b_n} e^{s_{a,n}/\tau}}  \right]
        \label{NL_loss}
    \end{equation}
    \normalsize
    
    It is important to note that both TL and NL are image recognition frameworks instead of image classification frameworks. For the purpose of TypoSwype, we opine that image recognition frameworks are more appropriate than classification because of the following.
    \begin{enumerate}[nosep]
        \item The desired list of domains (i.e. checking list) to which a cyber-defender may want to protect his network from phishing varies from company to company. For example, if a company is involved in the marine sector, his checking list of domains to which to check for typo-squatted variants may well be skewed toward the marine sector. Training a neural network to recognise a fixed list of domains via supervised learning will mean that the model needs to be retrained each time the checking list is updated.
        \item The number of domains that a cyber-defender may want to check for phishing may be very large. Training a neural network on large number of classes is especially challenging because of the softmax function's computational complexity at scale. \cite{hier_softmax}
    \end{enumerate}
  
\section{Methodology}
    \subsection{Conversion from string to swype-like images}
\begin{figure}[!htp]
        \centering
        \includegraphics[width=6cm]{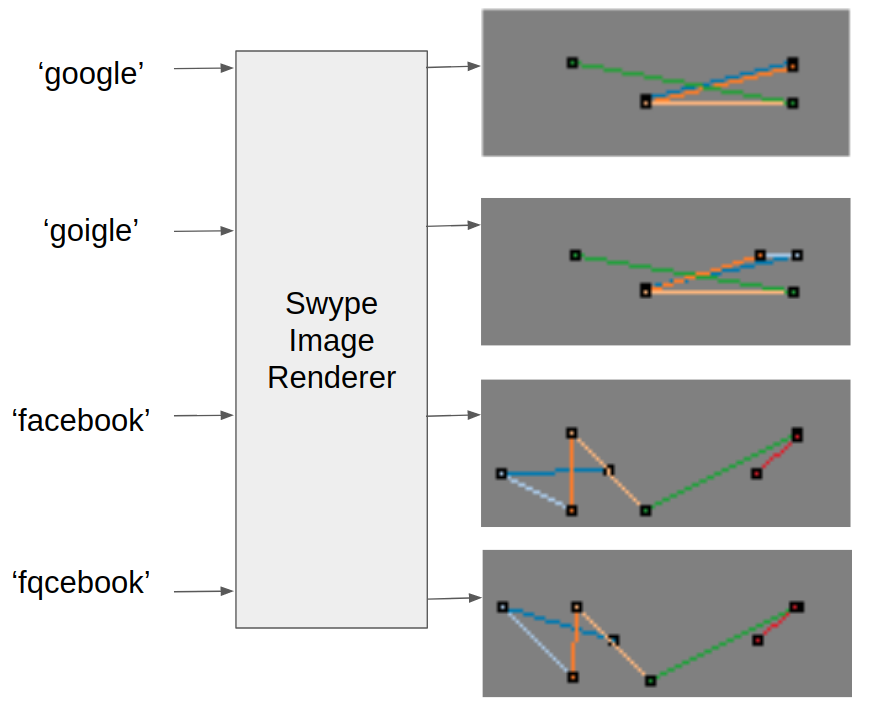}
        \caption{Conversion of input strings to swype-like images}
        \label{fig:swype_sample}
    \end{figure}

    The first step of our TypoSwype framework is to convert text to swype-like images. In this study, swype-like images were rendered based on the QWERTY keyboard layout as it is the most widely used keyboard format for English today. Each character in a string is then mapped to a grid location based on its physical location on the QWERTY keyboard. For example, the character "q" is mapped to grid $[1,0]$ as it is located on the second row and first column of the QWERTY keyboard. To ensure that lines are separated and not right on top of each other, a small amount of noise is added to each keyboard position corresponding to each character in the string. In our implementation, we added a random uniform noise between 0 and 0.1 for both axis. This corresponds to 10\% of a key on the keyboard as we defined each key on the keyboard as having a length and height of 1. Next, to take into account the sequence of characters in a string of text, a pre-set sequence of colors was used. For example, the first stroke between the $1^{st}$  and $2^{nd}$  character will always be blue, the next stroke between the $3^{rd}$ and $4^{th}$ character will always be light blue, etc. Finally, the corresponding location of the keys (together with noise) on the 4 $\times$ 10 grid, corresponding to the QWERTY keyboard, is multiplied by a factor of 10 and rendered into a 40 $\times$ 100 image via the Python Pillow package. Fig \ref{fig:swype_sample} illustrates 2 examples of popular domains and 2 possible typo-squatting variants. As can be seen, the swype-like images of the 2 domains and their possible typo-squatting variants are indeed similar, hinting the feasibility of this framework.
    
    \subsection{Neural Network design and training}
    Having rendered the strings into 40 $\times$ 100 swype-like images, the next step is to train the Encoder CNN model to output meaningful embeddings that can be used to detect typo-squatting domains. Our model is designed as shown in Table \ref{tab:cnn_archi}. The first 2 layers with same padding are important to ensure that information at the edges of the image are preserved. The outputs of the final layer are L2 normalised so that every embedding has a magnitude of 1. This is done so that both the TL and NL loss can be used to train the model. In particular, TL necessitates that the embeddings have an L2 norm of 1 as the Euclidean distance formulation of the TL loss would then be equivalent to minimising cosine similarity, since the squared Euclidean distance between normalized vectors is proportional to their cosine similarity \cite{l2normcos}.
    
    \begin{table}[!htbp]
    \caption{Encoder CNN Architecture}
    \label{tab:cnn_archi}
    \centerline{
        \begin{tabular}{|p{0.7cm}|p{0.7cm}|p{0.7cm}|p{1.5cm}|p{0.9cm}|p{1.3cm}|}
        \hline
        \textbf{Filters} & \textbf{Stride} & \textbf{Kernel} & \textbf{Convolution Type} & \textbf{Padding} & \textbf{activation} \\
        \hline
        8 & [1,1] & [3,3] & Conv2D & Same & leaky relu\\
        16 & [1,1] & [3,3] & Conv2D & Same & leaky relu\\
        64 & [1,1] & [3,5] & Conv2d & Valid & leaky relu \\
        64 & [1,2] & [3,5] & Conv2d & Valid & leaky relu\\
        128 & [2,2] & [3,5] & Conv2d & Valid & leaky relu \\
        128 & [2,2] & [3,5] & Conv2d & Valid & leaky relu\\
        128 & [2,2] & [3,5] & Conv2d & Valid & leaky relu\\
        1024 & - & - & Dense & - & tanh\\
        512 & - & - & Dense & - & leaky relu\\
        256 & - & - & Dense & - & l2 norm\\
        \hline
        \end{tabular}
        }
    \end{table}
    
    To train the model, we need a dataset. Due to the lack of any open source typo-squatting dataset, we made use of DNSTwist \cite{dnstwist} to generate samples of typo-squatting domains for the top 20k domains in the majestic million domains \cite{majestic1m}. In particular, DNSTwist permutates each of the 20k domain names to produce possible typo-squatting domains based on a predefined set of rules that cyber-attackers tend to use. During the process we noted that the domains produced by DNSTwist also takes into account keyboard distance by having an allowable dictionary for each key to be permuted. We also note that the dataset produced is biased toward small edit distances ($\leq2$). The dataset output by DNSTwist contains a total of approximately 2 million possible phishing domains, each derived from one of the top 20k domains of the majestic million domains.
    
    Next, we train the model via both TL and NL to determine their performances. For both models, we employed the following procedure to sample the negative samples.
    \begin{enumerate}[nosep]
        \item On the 1st step and after every 100 steps of training, we update the reference vectors. These vectors are the output vector of the Encoder (up till that point of training) after feeding in swype-like images of every item in the checking list as input. As we made use of the top 20k domains in the majestic million to create the dataset, the checking list also contains the same 20k domains.
        \item At each training step, we define the swype-like image of the typo-squatting domain and the label as the anchor and the positive respectively. To get the negative, we calculate the Euclidean distance between the anchor and each item in the checking list, excluding the label. Those with the highest similarity (i.e. smallest Euclidean distance) are the hard negative samples we will want to train on. Note that although NL loss makes use of cosine similarity as the similarity metric instead of Euclidean distance, the L2 normalization of the final layer of our Encoder ensures that the output vectors are L2 normalized. Comparing L2 Euclidean distance is proportional to comparing their cosine similarity \cite{l2normcos}.
    \end{enumerate}

    For NL, we took the top 8 hardest negative samples (i.e. $b_n = 8$) to train the Encoder while for TL, we sampled one of the top 8 negative samples at random to train the Encoder. We used a batch size of 64 to train both models and training was stopped after 1 epoch, due to the extremely large dataset and the fact that the model tends to converge before 1 epoch.
    
    As noted before, the dataset generated by DNSTwist takes into account keyboard distance and is biased toward small edit distances. As such, testing our model on a subset of this dataset and comparing it against the DLD algorithm as a baseline would be meaningless as the DLD algorithm would achieve perfect results. To determine if our model has learnt anything meaningful over DLD algorithm in terms of taking into account keyboard distance in its decision, we created a separate dataset, which we shall refer to as the test dataset, via the following steps:
    
    \begin{enumerate}[nosep]
        \item Take the top 100 domains from the majestic million dataset. For each domain, generate all possible domains, irrespective of keyboard distance, that are 1 LD edit distance away.
        \item For each generated domain, 
         \begin{enumerate}[nosep]
            \item If a deletion action was done, label that generated domain as a typo-squatting domain.
            \item If an insertion action was done, check if the inserted character is within 1 keyboard distance from the preceding and succeeding character. If the keyboard distance is equal to 1, label it as a typo-squatting domain; else if the keyboard distance is more than 3, label it as a benign domain; else drop the sample.
            \item If a substitution action was done, check if the substituted character is within 1 keyboard distance from the original character. If the keyboard distance is equal to 1, label it as a typo-squatting domain; else if the keyboard distance is more than 3, label it as a benign domain; else drop the sample.
        \end{enumerate}
    \end{enumerate}
    
    This test dataset will then be used to test our trained models against baseline DLD algorithms. Due to obvious reasons, the DLD algorithm will flag up every single item in this test dataset as a typo-squatting domain. We also note that the dataset is balanced, particularly, 45.5\% of the dataset, containing 8264 entries, was labelled as typo-squatting domains.
    
\section{Results and Discussion}
    There are 2 parts to the analysis. The first part pertains to the model's ability to detect typo-squatting. This is done by checking whether each domain in the test dataset, rendered as a swype-like image and fed to the model (i.e. Encoder) is similar (in terms of Euclidean distance) to any one of the swype-like images corresponding to domains in the checking list. If any of the Euclidean distance is less than 0.6 (chosen because the Euclidean distance between the anchor and hardest samples was found to converge to ~0.6 during training), we declare the domain as a possible typo-squatting domain. The model's performance in terms of detecting typo-squatting domains can then be determined by comparing the model's output against the labels generated while generating the test dataset.

    The second part pertains to the model's ability to correctly classify the typo-squatting domain for those that were labelled and also picked up by the model as typo-squatting domains. We define the matching domain as the domain in the checking list whose swype-like image's embedding (produced by the Encoder) has the smallest Euclidean distance to the embedding of the input domain's swype-like image. 
    
    \begin{table}[htbp]
    \caption{Accuracy \& F1-Score of various Experiments and Scenarios}
    \label{tab:accuracy_n_f1_score}
    \centerline{
        \begin{tabular}{|p{2cm}|c|c|c|c|}
        \hline
        \textbf{ } & \textbf{Experiment} & \textbf{DLD} & \textbf{TL} & \textbf{NL}\\
        \hline
        \multirow{3}{2cm}{Macro-F1 (Typo-squatting or not)} & 1 & 0.313 & 0.605 & 0.630 \\
                 & 2 & 0.313 & 0.617 & 0.628 \\ 
                 & 3 & 0.313 & 0.611 & 0.631 \\ 
        \hline
        Mean F1 & & 0.313 & 0.611 & 0.63 \\
        \hline
        \multirow{3}{2cm}{Accuracy (Domain Classification)} & 1 & 1.00 & 0.997 & 0.998 \\
                 & 2 & 1.00 & 0.997 & 0.997 \\
                 & 3 & 1.00 & 0.996 & 0.998 \\
        \hline
        Mean Accuracy & & 1.00 & 0.997 & 0.998 \\
        \hline
        \end{tabular}
        }
    \end{table}
    
\begin{figure}[!htp]
        \centering
        \includegraphics[width=9cm]{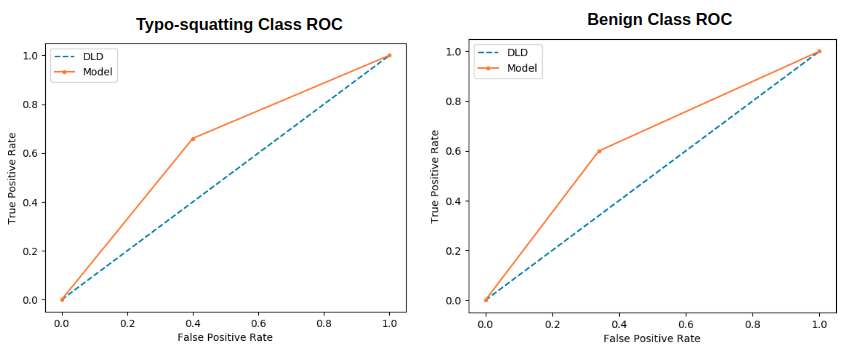}
        \caption{ROC Curves of TypoSwype-NL}
        \label{fig:roc_curve}
    \end{figure}

    The results of our TypoSwype method using both TL and NL are illustrated in Table \ref{tab:accuracy_n_f1_score}. As can be seen, our TypoSwype framework, whether being trained by TL or NL, results in a higher F1 score in detecting typo-squatting domains that are 1 edit distance away from the original domain as compared to DLD. Furthermore, the domain classification accuracy is similar across the 3 methods. These clearly demonstrates TypoSwype's ability to detect similarity among swype-like images for typo-squatting detection. It should also be noted that despite training the Encoder on a dataset generated by DNSTwist and subsequently testing on a separately generated dataset, which implies an underlying distributional shift in the input data, we were still able to achieve almost a one time improvement in F1 score over the conventionally used DLD algorithm for typo-squatting detection. Fig \ref{fig:roc_curve} also shows the Receiver Operating Curve (ROC) for the best performing NL model in relation to the baseline DLD algorithm on the test dataset. Clearly, our proposed method is advantageous over the baseline DLD algorithm in taking account keyboard distance into consideration when detecting typo-squatting domains.
    
    \begin{figure}[!htp]
        \centering
        \includegraphics[width=7cm]{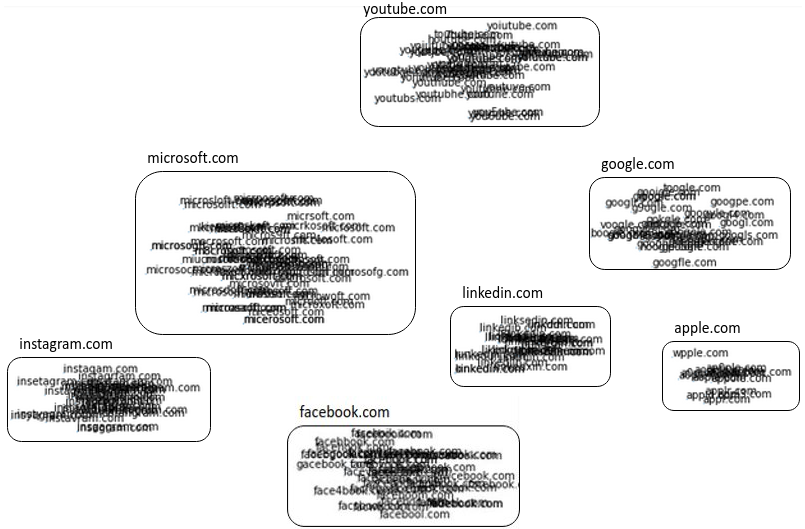}
        \caption{Clustered Typos of popular domains}
        \label{fig:typo_clusters}
    \end{figure}
    
    Finally, as shown in Fig \ref{fig:typo_clusters}, typo-squatting variants of the various depicted domains can be easily clustered together. This not only showcases the Encoder's ability to produce meaningful embeddings of the input swype-like images. It also shows that the TypoSwype framework of rendering strings as swype-like images is able to preserve sufficient amount of information for detecting and classifying typo-squatting variants, making it a candidate for other spell-check disciplines.

\section{Conclusion and Future Work}
    We have demonstrated the possibility of using Swype-like images together with SOTA image recognition algorithms to conduct typo-squatting detection. This approach precludes the need for carefully tuning the weights, as in the case of the weighted DLD approach proposed by \citeauthor{Samuelsson}, which typically results in over fitting to a particular dataset \cite{Samuelsson}. This over fitting problem could be the reason why the algorithm together with a common set of weights has not seen widespread adoption unlike the DLD algorithm. Furthermore, it should also be noted that the technique proposed here allows batch processing of multiple input domains at one shot, facilitating faster run time compared to DLD approaches which operate on a single input at a time. However, a curated dataset is needed to train and test the model. Future work could focus on ways to create a standard dataset and apply such algorithms to other domains that require spell checks. 

\bibliographystyle{unsrtnat}
\bibliography{references}

\end{document}